\documentstyle[epsf]{elsart}

\begin{document}

\begin{frontmatter}
\title{{\bf The Anisotropy of Cosmic Ray \\
Arrival Directions around 10$\bf ^{18}$eV}}

\thanks[EK]{
Present address: Institute for Cosmic Ray Research,
Univ. of Tokyo,\newline
Midori-cho 3-2-1, Tanashi, Tokyo 188-8502, Japan \newline
e-mail: erina@icrr.u-tokyo.ac.jp\newline
Tel +81 424 69 9595 / Fax +81 424 62 3096
}
\author{N. Hayashida, M. Nagano, D. Nishikawa, H. Ohoka, N. Sakaki},
\author{M. Sasaki, M. Takeda, M. Teshima, R. Torii, T.Yamamoto, S. Yoshida},
\address{Institute for Cosmic Ray Research, University of Tokyo, Tokyo 188-8502, Japan}
\author{K. Honda},
\address{Faculty of Engineering, Yamanashi University, Kofu 400-8511, Japan}
\author{N. Kawasumi, I. Tsushima},
\address{Faculty of Education, Yamanashi University, Kofu 400-8510, Japan}
\author{N. Inoue}, 
\author{E. Kusano\thanksref{EK}},
\author{K. Shinozaki, N. Souma},
\address{Department of Physics, Saitama University, Urawa 338-8570, Japan}
\author{K. Kadota, F. Kakimoto},
\address{Department of Physics, Tokyo Institute of Technology,Tokyo 152-8551, Japan}
\author{K. Kamata},
\address{Nishina Memorial Foundation, Komagome, Tokyo 113, Japan}
\author{S. Kawaguchi},
\address{Faculty of General Education, Hirosaki University, Hirosaki 036-8560, Japan}
\author{Y. Kawasaki},
\address{Department of Physics, Osaka City University, Osaka 558-8585, Japan}
\author{H. Kitamura},
\address{Department of Physics, Kobe University, Kobe 657-8501, Japan}
\author{Y. Matsubara},
\address{Solar-Terrestrial Environment Laboratory, Nagoya University, Nagoya 464-8601, Japan}
\author{K. Murakami},
\address{Nagoya University of Foreign Studies,
 Nissin, Aichi 470-0197, Japan}
\author{Y. Uchihori},
\address{National Institute of Radiological Sciences, Chiba 263-8555, Japan}
\author{H. Yoshii}
\address{Department of Physics, Ehime University, Matsuyama 790-8577, Japan}

\begin{abstract}
Anisotropy in the arrival directions of cosmic rays
with energies above 10$^{17}$eV is studied
using data from the Akeno 20 km$^2$ array and the Akeno Giant Air Shower Array (AGASA), 
using a total of about 114,000 showers observed over 11 years.
In the first harmonic analysis, we have found strong anisotropy
of $\sim$ 4\% around 10$^{18}$eV, corresponding to a chance probability
of $\sim 0.2\%$ after taking the number of independent trials into account.
With two dimensional analysis in right ascension and declination,
this anisotropy is interpreted as an excess of showers near
the directions of the Galactic Center and the Cygnus region.
\end{abstract}
\end{frontmatter}

\newpage

\section{Introduction}
Searches for anisotropy
in the arrival directions of high energy cosmic rays 
have been made by many experiments so far and 
the arrival direction distribution
of cosmic rays is found to be quite  isotropic
over a broad energy range.  In most  experiments 
 harmonic analysis in right ascension (RA) has been applied
 to find a wide-angle cosmic ray flow. 
Results  of various experiments up to 1980 are summarized
in Linsley and Watson \cite{linsley81}, up to 1983  in 
Lloyd-Evans and Watson \cite{evans83} ,
 and to 1991 in Watson \cite{watson92} . 
Above 10$^{15}$eV, most amplitudes published so far are
upper limits and increase with energy as E$^{0.5}$ up to 
a few $\times$10$^{19}$eV, 
due to the number of events decreasing with energy
like E$^{-2}$.

One possibly significant signal of wide-angle cosmic ray flow
 claimed so far is from  the
Haverah Park experiment in the energy region near 10$^{17}$eV,
which shows an amplitude (1.52 $\pm$ 0.44)\% at RA = 212$^{\circ}$
 $\pm$ 17$^{\circ}$  
with chance probability of 0.3\%  \cite{evans83}. 
The interpretation of significant anisotropy in a narrow
energy range may be very difficult and 
their result should be confirmed with much larger statistics.  
The Yakutsk group claimed a significant anisotropy
with an amplitude (1.2 $\pm$ 0.3)\% 
and chance probability  0.17\% in the energy
region 10$^{16.5}$ $\sim$ 10$^{17.5}$eV \cite{efremov96}.
The phase, however, is 124$^{\circ}$ in the right ascension so the direction
is 90$^{\circ}$ different from the Haverah Park result.

The Haverah Park group  also claimed 
evidence for an enhancement from southern latitudes in the 
range 5$\times$ 10$^{17}$eV to 10$^{19}$eV and for the tendency of
primary cosmic rays to arrive from high northern galactic latitude above
10$^{19}$eV \cite{eames85}.

The Fly's Eye group investigated the anisotropy by dividing the sky
into six lobes of equal solid angle and comparing the number of
detected cosmic rays in each lobe with the number expected from an
isotropic intensity \cite{cassiday90}.
  The lobes investigated are the directions of
the north Galactic pole, the south Galactic pole, the center of the Galaxy, 
the anti-center of the Galaxy, forward along the solar revolution
(the Sun's orbit about the Galaxy's center) and
backward along that revolution.  They could not find any statistically
significant anisotropy above 10$^{17}$eV, but
detected some excess from the Galactic north sky lobe above 
10$^{19.5}$eV consistent with the Haverah Park indication of a northern
excess at the highest energies.

In this report we present the anisotropy in arrival directions
of cosmic rays around 10$^{18}$eV observed by AGASA. That beyond 10$^{19}$eV
will be reported in a separate paper.

\section{Experiment}

The Akeno Giant Air Shower Array (AGASA) consists of 111 
scintillation detectors of 2.2 m$^{2}$ area each,
which are arranged with inter-detector spacing of about 1 km
over 100 km$^2$ area.
Akeno is located at latitude 35$^{\circ}$ 47'$N$
and longitude 138$^{\circ}$ 30'$E$ at an
average altitude of 900 m above sea level.
Details of the AGASA array are described in Chiba et al. \cite{chiba92}.

The whole area is divided into 4 branches, 
called the Akeno Branch (AB),
the Sudama Branch (SB), the Takane Branch (TB), and the Nagasaka Branch (NB).
The data acquisition started independently in each branch,
and the four branches were unified in December of 1995.
The present result includes data up to July 1995 before the
unification.  Data from the 20 km$^2$ array \cite{teshima86}
are included in this analysis.  Data acquisition from that array
started in 1984, and it became part of the AB branch of AGASA 
in February 1990.
The triggering requirement is a coincidence of more than five adjacent 
detectors, each with a signal greater than 20\% of that produced by
a muon traversing vertically the scintillator of 5 cm
thickness.
About 99\% of triggered events are accidentally coincident and only
   1\% are real air showers, which are selected in the procedures of
 fitting the particle densities of all detectors within 2.5 km from the core
to the empirical lateral
distribution and their arrival times to the empirical shower
front structure. 
About 230,000 events are identified as extensive air showers
over the total observation period of 11 years. 
The typical angular resolution is 3 degrees and 1.5 degrees 
for 10$^{18}$eV and 10$^{19}$eV showers, respectively.

\section{Results}

The data sets used in the present analysis are listed in Table 1. 
The column marked Sel.1 (Selection 1) in Table 1 shows 
the number of events selected based on the following 
conditions: 
the core is inside the array, the number of hit detectors is $\geq 6$,
and the reduced $\chi^{2}$ in determining the arrival direction
and the core position is less than 5.0.
All events with zenith angles $\leq 60^{\circ}$ are used 
in the present analysis.
About 114,000 events remain after this Selection 1.

One of the conventional methods to search for any global anisotropy in 
the arrival directions of cosmic rays is to apply harmonic analysis
to the right ascension distribution of events.
That is, the method is to fit the distribution to a sine wave with 
period $2\pi/m$ ({\it m}-th harmonic)
to determine the amplitude and 
phase of the anisotropy.  The {\it m}-th harmonic amplitude, {\it r}, and 
phase of maximum, \(\theta\), are obtained for a sample of {\it N}
measurements of phase, \(\phi_{1},\phi_{2},\cdots,\phi_{n}
(0\leq\phi_{i}\leq2\pi)\) from:

\begin{equation}
 r = (a^{2}+b^{2})^{\frac{1}{2}}
\end{equation}
\begin{equation}
\theta=\tan^{-1}(b/a)
\end{equation}
where,
$a=\frac{2}{n}\sum_{i=1}^{n}\cos m\phi_{i},\ 
b=\frac{2}{n}\sum_{i=1}^{n}\sin m\phi_{i}$.

The following $k$ represents the statistical significance.
If events with total number {\it N} are uniformly distributed in right 
ascension, the chance probability of observing the 
amplitude $\geq$ $r$ is given by,
\begin{equation}
 P = \exp(-k),
\end{equation}
where 
\begin{equation}
 k = Nr^{2}/4.
\end{equation}
Results of first harmonic analysis in right ascension
using the events after Selection 1
are shown in Figure 1.
The amplitude (top), the phase (middle), and the significance $k$ (bottom) 
are shown as a function of primary energy threshold.
Each point is obtained by summing over events with more than the corresponding 
energy.
Clearly, $k$ $\sim$ 10 around 10$^{18}$eV is surprisingly high, corresponding
to a chance probability of 0.005\%. 
We have searched for the energy bin width which gives the maximum $k$
-value, and find that the region 10$^{17.9}$eV - 10$^{18.3}$eV 
gives the maximum 
$k$-value of 11.1. This means the showers which contribute to 
the anisotropy are distributed in the energy range of 0.4 decade.
In Figure 2, the right ascension distributions of events are shown 
in the energy ranges $<$10$^{17.9}$eV (top), 
10$^{17.9}$eV $\sim$ 10$^{18.3}$eV (middle),
and $>$10$^{18.3}$eV (bottom).
A clear excess is found  around 300$^{\circ}$ in the right ascension 
distribution of events in the energy region 
10$^{17.9}$eV $\sim$ 10$^{18.3}$eV (middle) and is not found
in the other energy ranges (top and bottom). 

In Table 2, the results of harmonic analysis 
are listed as a function of threshold energy in each 0.5 decade.
We also listed the results in the differential bins with energy
ranges of a factor of two from $1/8EeV$ to 8EeV in Table 3,
for the comparison with the world data.
According to these tables, the chance probability is estimated to be
$\sim 0.21\%$ by taking the number of independent trials into account.

In searching for anisotropy,
rates from different regions on the celestial sphere are compared.
Therefore uniform observation time in right ascension
is quite important in this analysis.
There are various effects which can produce spurious anisotropies,
such as a temporary detector inefficiency, or communication
trouble, or spurious events due to lightning, or change of
observed rates due to temperature and pressure variation.

In the following we try to exclude data that might include 
spurious events by checking the data set in each day and each branch.

\begin{enumerate}
\item If the detection efficiency of each day were constant and 
there was no lack of observing during any day, 
the daily number of events should follow a Gaussian 
 distribution centered on the average value.
We selected only those days for which the number
of events is  within $\pm 2\sigma$ of the average. 

\item If the event distribution were random in each day,
the distribution of daily $k$-values (Eq. (3)) should follow $\exp(-k)$
(Rayleigh test). Days which have $k$ greater than 2 are excluded.

\item To find days which include a sudden increase or decrease of events for
a short time, we have applied the Kolmogorov-Smirnov test (K-S test) 
on the data for each day and selected those days having a maximum deviation 
less than 90\% of the boundary.
\end{enumerate}

With these three criteria, bad days that could cause
spurious anisotropy in arrival directions were excluded.
The numbers of remaining events after these selections
are listed in the column marked Sel.2 (Selection 2) in Table 1.
About 70\% of the events were selected.

Using the data set after Sel.2, the first harmonic 
analysis has been done and the results are shown in Figure 3. 
We can still see clear peaks in the $k$-plot,
$k$ $\sim$ 7 around 10$^{18}$eV, corresponding to a chance probability 
of 0.06\%. The decrease of the value $k$ can be explained
by the decrease in the number of events (70\%).
It should be noted that the anisotropy amplitude and phase did not change 
after these selections. This means the observed anisotropy can not be 
due to those spurious causes. We have also carried out 
harmonic analysis on the cut data (30\%) and they also show a
Rayleigh power with $k$ $\sim$ 3 at 10$^{18}$eV. 
Considering the difference in the number of selected and cut
events, it is concluded that the significance of the observed anisotropy 
is independent of the above selections.

Any spurious variation would likely arise from diurnal variations
(temperature and barometric pressure effects) and should be more
evident in solar time than in sidereal time.
We checked the solar time variation of the number of air shower events 
in Sel.2 using harmonic analysis.
That is, the analysis was done using the arrival time of each event in solar time
instead of sidereal time.
 Around 10$^{18}$eV the amplitude is about 1\% in solar time
and $k$ is less than 1 as shown in Figure 3 by thin dotted points.
At Akeno the amplitudes of first harmonic and second harmonic
pressure variation are about 0.5 mb and 0.9 mb in solar time 
at 3 hour and 9 hour, respectively, throughout a year.
The expected amplitude due to the pressure variation is 
smaller than 0.4\% in solar time \cite{murakami}.
Even if there were significant anisotropy of 1\% in the solar time
due to other reasons, the amplitude due to the daily variation
must be reduced considerably when analyzed in sidereal time.
Conversely, if the 4\% anisotropy in right
ascension seen in the present experiment were due to a daily
variation, then the amplitude should be larger 
in the solar time analysis.
We can conclude that the observed anisotropy 
in sidereal time is not due to a solar effect. 
These considerations indicate that the observed anisotropy is genuine.

In Figures 4 and 5, the arrival direction distributions in equatorial
coordinates are shown.
They show the ratio of the number of observed events to the number 
expected and the statistical significance of the deviations, respectively.
Here, the energy region of 10$^{17.9}$ $\sim$ 10$^{18.3}$eV is selected 
to maximize
the harmonic analysis $k$-value.
Since the geographical latitude of Akeno observatory is 35$^\circ$ 47'$N$,
we can not observe events with declination less than -25$^{\circ}$,
as long as we use showers with zenith angles less than 60$^{\circ}$.

The number of expected events at each right ascension and declination is 
estimated as follows.
The sky is divided into declination bands (width of 1 degree), and
the number of events in each declination band is calculated
($f(\delta) d \delta$).
Since non-uniformity of the observation time in right ascension
is less than 1\% from the data, we estimate
the expected event density as  $g(\alpha,\delta) = f(\delta)/360$ in
each right ascension and declination bin by assuming 
constancy in right ascension.  
In these figures, 
we have chosen a circle of 20$^{\circ}$ radius to evaluate
the excess. We have integrated the expected event density
inside this circle $\int_s g(\alpha,\delta) d\alpha d\delta$ and then compared
with the observed number.
We have examined with four different radii of 10$^\circ$, 15$^\circ$, 
20$^\circ$, and 30$^\circ$ centered near the Galactic center 
and obtained significances, $ 2.6 \sigma$, $ 2.7 \sigma$, $ 4.1 \sigma$ 
and $ 2.8 \sigma$, respectively.
The radius of 20$^\circ$ gave the maximum deviations.

In the significance map with beam size of $20^\circ$, 
a $4 \sigma$ excess (obs./exp. = 308/242.5)
can be seen near the direction of the Galactic Center.
In contrast, near the direction of anti-Galactic Center
we can see a deficit in the cosmic ray intensity ($ -3.7\sigma$).
An event excess from the direction of the Cygnus region 
is also seen in the significance map at the 3 sigma level.

\section{Discussion}

An anisotropy of amplitude 4\% around 10$^{18}$eV was found
in first harmonic analysis. 
With a two dimensional map, we can identify this
as being due to event excesses of $4\sigma$ and $3\sigma$ 
near the galactic center and the Cygnus region, respectively.
The observed anisotropy seems to be correlated with the galactic structure.

As shown in Table 4,  such anisotropy has not been observed by
previous experiments.  Since the latitudes of the Haverah Park and Yakutsk
are around 60 degrees, the direction of significant excess in
the present experiment near the galactic center can
not be observed by those experiments and hence a significant
amplitude in harmonic analysis might be absent in their data.
However, the possible enhancement at
southern galactic latitudes in 5$\times$10$^{17}$ $\sim$ 10$^{19}$eV
claimed by the Haverah Park experiment may be related to the
present experiment.     
Statistics from the Fly's Eye experiment do not appear to be
sufficient to support or refute the present result.

One possible explanation of the anisotropy reported here involves
cosmic ray protons.
In Figure 6, a schematic view of the galactic spiral structure is shown
\cite{georgelin88}.
The observed regions of excess are directed toward the galactic plane.
Their directions are shown by the hatched region in the figure and seem to
be correlated with the nearby spiral arms.
The Larmor radius of a proton with energy 
10$^{18}$eV is estimated to be $\sim$ 300 pc in our galaxy, 
which is comparable with the scale height of the Galaxy's magnetic field.
Near this energy the slope of the cosmic ray energy spectrum changes.
It becomes steeper 
above 10$^{18}$eV
 \cite{nagano92,yoshida95} as
 the leakage of cosmic rays from our galaxy seems to become
more rapid than at lower energies.
In the leaky box model, the amplitude of anisotropy and 
the energy spectra at the production site and 
observation site are closely related \cite{hillas84}.
If we denote the cosmic ray residence time by $\tau$(E), the amplitude of
the anisotropy is proportional to $1/\tau$(E).
That is, if the observed energy spectrum becomes steeper at around
 10$^{18}$eV, the anisotropy should become stronger at that energy.
However, the direction of anisotropy need not point toward the
nearby galactic arm, since scattering is diffusive in the leaky box model.
According to the Monte Carlo simulation by Lee and Clay 
\cite{lee95},
a proton anisotropy of 10\% $\sim$ 20\% amplitude
is expected at RA $\sim$ 300$^{\circ}$
using an axisymmetric concentric ring model
of the galactic magnetic field with interstellar turbulence
of a Kolmogorov spectrum.  The source 
 distribution is assumed to be uniform within the galactic disk
and both a non-random and turbulent magnetic halo with
various field strengths are taken into account.
If the observed anisotropy is due to protons, we can estimate 
the proton abundance as to be about 20\% $\sim$ 40\% of  
all cosmic rays, by comparing our result of 4\% amplitude
 with their simulation.

Another possible explanation is that the anisotropy is due to neutron 
primary particles.
Neutrons of 10$^{18}$eV have a gamma factor of 10$^9$ and
their decay length is about  10 kpc. Therefore they can propagate
from the galactic center without decaying or bending
by the magnetic field. 
In the cosmic ray acceleration regions, there may be ambient photons 
or gases. The accelerated heavy nuclei should interact 
with these photons or matter, and spill out neutrons. 
The acceleration region may have enough size and
magnetic field strength to confine the charged particles, 
while the produced neutrons can escape easily
from the site. In this scenario, the heavy dominant 
chemical composition below 10$^{18}$eV \cite{gaisser93} and the lack of 
anisotropy below 10$^{17.9}$eV (due to the short
neutron lifetime) can be naturally explained.
Below 10$^{18}$eV, the neutron energy spectrum strongly depends on the 
source distance. On the other hand, above 10$^{18}$eV, it depends 
on the source energy spectrum. We have tried to fit the observed $k$
distribution with the expected one obtained by assuming the energy  
spectrum and the source distance, 
however, we found the neutron energy spectrum with the power law spectrum 
of $E^{-2}-E^{-2.5}$ which can be expected from the acceleration model 
does not agree with the $k$ distribution above 10$^{18.5}$eV.
We need to use a steeper energy spectrum with an index of -3 $\sim$ -4, or
we need to introduce the cutoff in the energy spectrum at 10$^{18.5}$eV.
For example, we could fit well the observed $k$ distribution with the 
reasonable parameters $\gamma$ = -2.5, $D$ = 10 kpc, 
$E_{cut}$ = 10$^{18.5}$eV as shown 
in Fig.7. Where $\gamma$ is a index of the differential energy spectrum.
The cutoff energy of 10$^{18.5}$eV or steeper energy spectrum
may be natural, if we consider the maximum energy of galactic cosmic rays.

In this section, we have discussed two possibilities of the anisotropy origin; 
however, it is difficult to interpret the present data.
More accumulation of the data, observation in the southern hemisphere, 
and the determination of energy spectrum in the excess region are important to
confirm the experimental result and to discriminate
two possibilities.

\section*{Acknowledgment}
We are grateful to Akeno-mura, Nirasaki-shi, Sudama-cho, Nagasaka-cho,
Takane-cho and Ohoizumi-mura for their kind cooperation.
The authors also wish to acknowledge the valuable help by other members
of the Akeno Group in the maintenance of the array.
We would like to thank Dr. P. Sommers and Dr. S. F. Taylor for the 
valuable discussions.

\newpage

\begin{table}
\caption{The data sets used in the present analysis. 
``Akeno I'' is a data set taken by the 20km$^2$ array from
1984 to 1990.  Sel. 1 is obtained after the usual AGASA data selection. 
Sel. 2 is after the good day cut.}
\centerline{
\begin{tabular}{lcrrrr} \hline 
Branch    & period        & period(yr) &  Showers & Sel. 1    & Sel. 2 \\ \hline
Akeno I    & 840909-900216 & 3.5        &   12323    &    9310   &    6857  \\ 
Akeno II   & 900217-930419 & 3.2        &   34493    &   14153   &    9283  \\
Akeno III  & 930623-941215 & 1.7        &   26432    &   12304   &    8348  \\ 
Akeno-Sudama & 941226-950630 & 0.7      &   18681    &   11827   &    7517  \\ 
Nagasaka I & 910308-930709 & 2.4        &   21379    &    9581   &    6037  \\
Nagasaka II& 930811-950630 & 1.10       &   18670    &    8862   &    6031  \\ 
Sudama I   & 900731-930512 & 2.10       &   27769    &   11858   &    7852  \\
Sudama II  & 930821-941214 & 1.5        &   15408    &    7129   &    5033  \\
Takane I   & 901115-930914 & 2.10       &   34882    &   17149   &   11762  \\
Takane II  & 930924-950630 & 1.9        &   22640    &   11811   &    8418  \\ 
\hline 
total     &               &             &  232677    &  113984   &   77138  \\
\hline
\end{tabular}
}
\end{table}

\begin{table}
\caption{The results of the first and second harmonic 
in right ascension as a function of energy.}
\centerline{
\begin{tabular}{lrrrrl} \hline 
         & Log(energy) & number of events & Amplitude[\%] &  Phase & P$_{prob}$  \\ 
\hline
         & $\ge$17.5eV & 81904 & 1.4 & 301 & 0.014 \\
First    & $\ge$18.0eV & 27600 & 3.6 & 293 & 0.00009 \\
harmonic & $\ge$18.5eV &  4096 & 3.5 & 268 & 0.26 \\
         & $\ge$19.0eV &   495 & 4.4 & 28 & 0.77 \\ \hline 
         & $\ge$17.5eV & 81904 & 0.4 & 207 & 0.67 \\
Second   & $\ge$18.0eV & 27600 & 0.6 & 215 & 0.71 \\
harmonic & $\ge$18.5eV &  4096 & 1.4 &  7 & 0.80 \\
         & $\ge$19.0eV &   495 & 12 & 109 & 0.13 \\ \hline 
\end{tabular}
}
\end{table}

\begin{table}
\caption{The differential results of the first harmonic analysis 
in right ascension as a function of energy.}
\centerline{
\begin{tabular}{lcrrrll}\hline 
Bin & Energy Range/EeV &  \#   & Amplitude[\%] & Phase & k    & P$_{prob}$ \\
\hline
E2  &  1/8 - 1/4       & 19146 & 1.6     & 211   & 1.37 & 0.25	\\
E3  &  1/4 - 1/2       & 32921 & 1.2     &  35   & 1.32 & 0.26	\\
E4  &  1/2 - 1.0       & 31657 & 1.0     & 298   & 0.87 & 0.41	\\
E5  &  1.0 - 2.0       & 18274 & 4.1     & 300   & 7.95 & 0.00035 \\
E6  &  2.0 - 4.0       &  6691 & 3.1     & 269   & 1.62 & 0.19	\\
E7  &  4.0 - 8.0       &  1913 & 2.9     & 278   & 0.41 & 0.66	\\
\hline
\end{tabular}
}
\end{table}

\begin{table}
\caption{Comparison of harmonic analysis 
in right ascension by various experiments in the energy region around
$10^{18}$eV.}
\centerline{
\begin{tabular}{lrrrlrllr} \hline 
Experiment & latitude & energy & \# & Amplitude[\%] &  Phase & $k$ & ref. \\ \hline
present &  35.47$^{\circ}$N & 1-2 EeV & 18274 & 4.1$\pm$1.0 & 300 & 7.9 \\
Haverah Park &  53.58$^{\circ}$N & 1-2 EeV & 7320 & 2.1$\pm$1.7 & 70 & 0.80 & \cite{gillman93}  \\
Yakutsk & 61.7$^{\circ}$N & 1-1.8 EeV & 14972 & 1.6$\pm$1.2 & 198 & 0.97 &
\cite{efremov96} \\
Fly's Eye & 40.2$^{\circ}$N & 1-2 EeV & 1579 & 6.6 & 318 & 0.09 & 
 \cite{cassiday90}\\ 
\hline 
\end{tabular}
}
\end{table}

\newpage

\begin{figure}
\end{figure}

\begin{figure}
\caption{Right Ascension distribution in the energy range of $<$10$^{17.9}$eV
(top), 10$^{17.9}$-10$^{18.3}$eV (middle), $>$10$^{18.3}$eV (bottom)}.
\end{figure}

\begin{figure}
\caption{The result of the first harmonic analysis in right ascension 
using data after Sel.2.
The amplitude (top), phase (middle), $k$ (bottom) of anisotropy 
in each energy bin
are plotted as a function of the primary energy.
The result of the first harmonic analysis in solar time applied to the
same data sets is drawn by thin dotted points.}
\end{figure}


\begin{figure}
\caption{Map of ratio of the number of observed events to expected ones
in equatorial coordinate. Events within a radius of 20$^{\circ}$ 
are summed up in each bin. Solid line shows Galactic Plane.
 G.C. marks the galactic center}
\end{figure}

\begin{figure}
\caption{Significance map of excess or deficit events.
Events within radius of 20$^{\circ}$ are summed up in 
each bin.}
\end{figure}

\begin{figure}
\caption{The spiral structure of our Galaxy [10].
The shaded regions correspond to the excess directions of the
present experiment.}
\end{figure}

\begin{figure}
\caption{The experimantal result can be well explained by the 
neutron source model with parameter, $\gamma$ = -2.5, 
$D$ = 10 kpc, $E_{cut}$ = 10$^{18.5}$eV.}
\end{figure}

\end{document}